\documentstyle[11pt,newpasp,twoside,psfig]{article}
\index{instructions}
\index{guidelines}

\def\edcomment#1{\iffalse\marginpar{\raggedright\sl#1\/}\else\relax\fi}
\reversemarginpar

\begin{document}
\title{Neutron Star Formation and Birth Properties}
 \author{Hans-Thomas Janka}
\affil{Max-Planck-Institut f\"ur Astrophysik, Karl-Schwarzschild-Str.\ 1,
D-85741 Garching, Germany}

{\small
\begin{abstract}
Our current knowledge of neutron star formation, progenitors,
and natal masses, spins, magnetic fields, and space velocities 
is briefly reviewed from a theorist's perspective. More
observational information is badly needed to constrain 
theoretical possibilities.
\end{abstract}
}

\section{Introduction: The beginning matters}

Although only a wink in its life, the moment of birth of a 
neutron star marks a spectacular astrophysical event
with far-reaching consequences. Neutron stars originate from
the apocalyptic death of massive stars in supernova (SN) explosions.
While Baade and Zwicky (1934) first came up with this visionary
suggestion, the link is now unambiguously established
by associations of pulsars and compact X-ray sources with young 
gaseous SN remnants, e.g.\ in the famous cases of the
Crab pulsar and Crab nebula, Vela pulsar and nebula, or the 
Cassiopeia A remnant with the compact central object that
was pinpointed with high resolution by the {\em Chandra} 
X-ray Observatory. 

Neutron stars certainly belong to the most exotic known objects.
With the size of roughly three gravitational radii they contain 
more than a solar mass
of matter at a density exceeding that in atomic nuclei. The
gravitationally bound object is kept in mechanical equilibrium
by repulsive interactions and degeneracy pressure of nucleons
that balance the enormous pull of gravity. The extreme compactness
allows pulsars to rotate with periods as low
as a millisecond and to possess surface magnetic fields up to
15 orders of magnitude higher than that of the Earth.
Extraordinary conditions like these make them unique astrophysical
laboratories for nuclear physics, particle physics, and  
gravitational physics.

\begin{figure}[htp!]
\centerline{\psfig{file=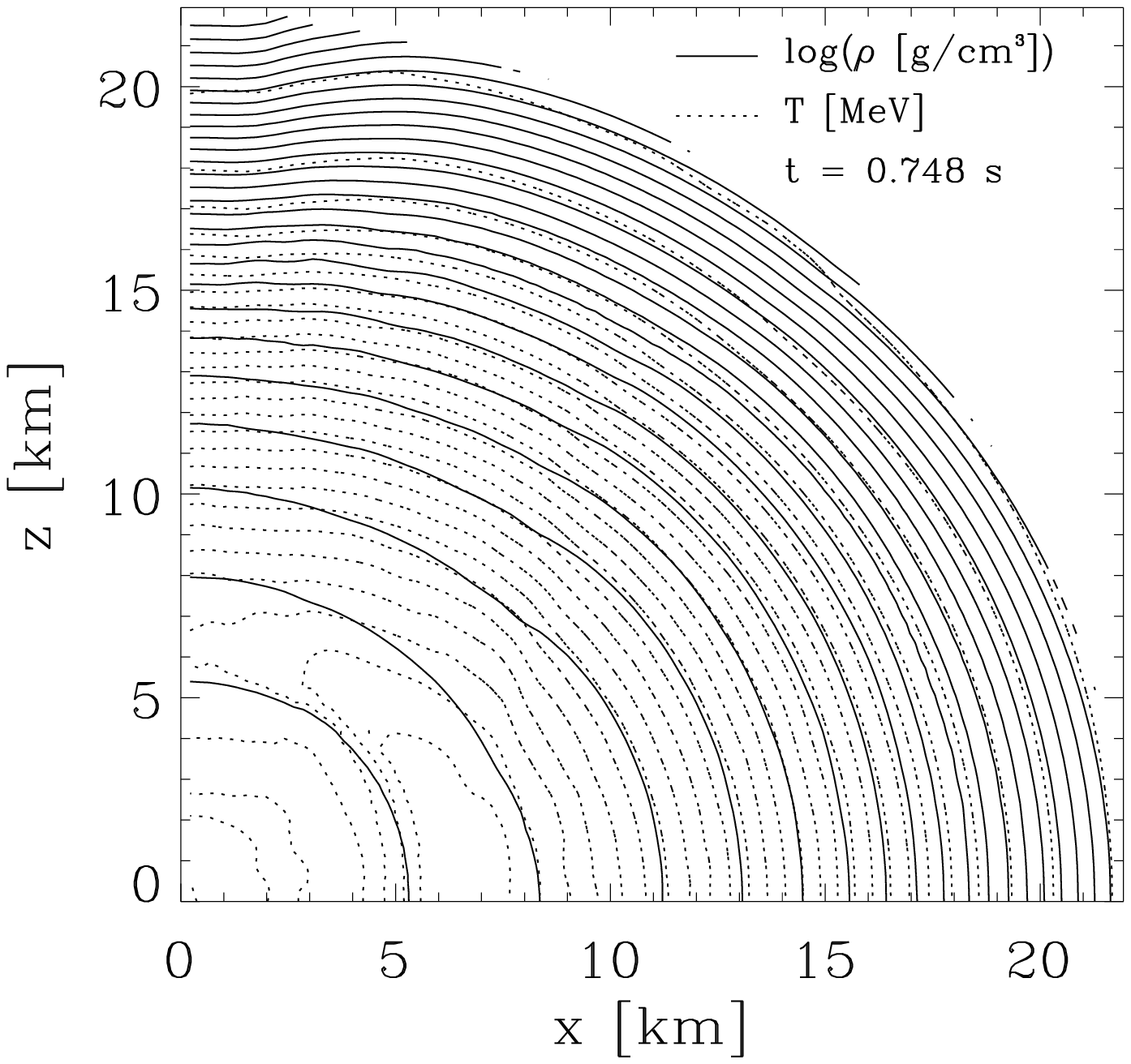,width=0.49\textwidth}
            \psfig{file=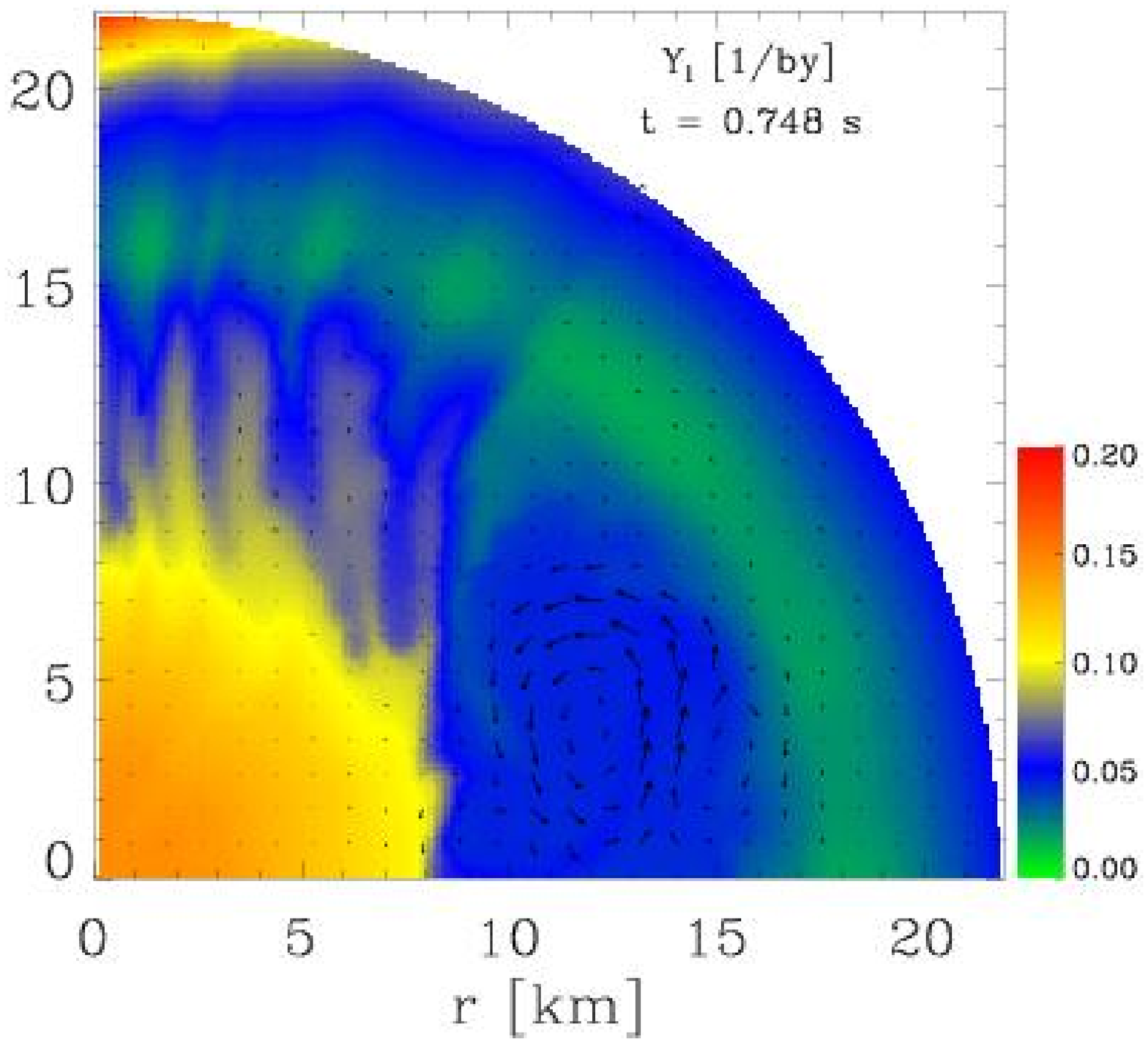,width=0.49\textwidth}} 
\vspace{7pt}
\centerline{\psfig{file=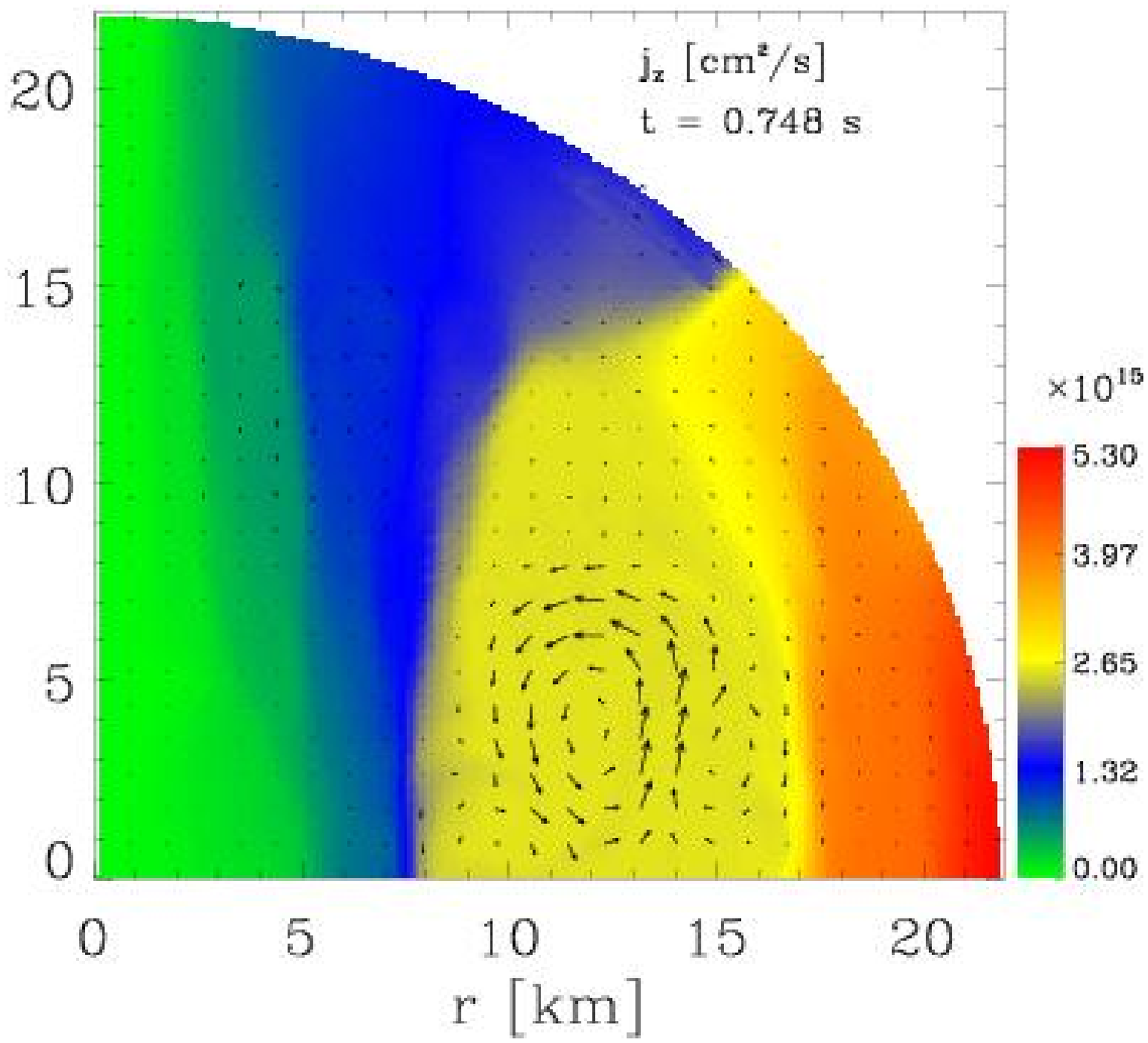,width=0.49\textwidth}  
            \psfig{file=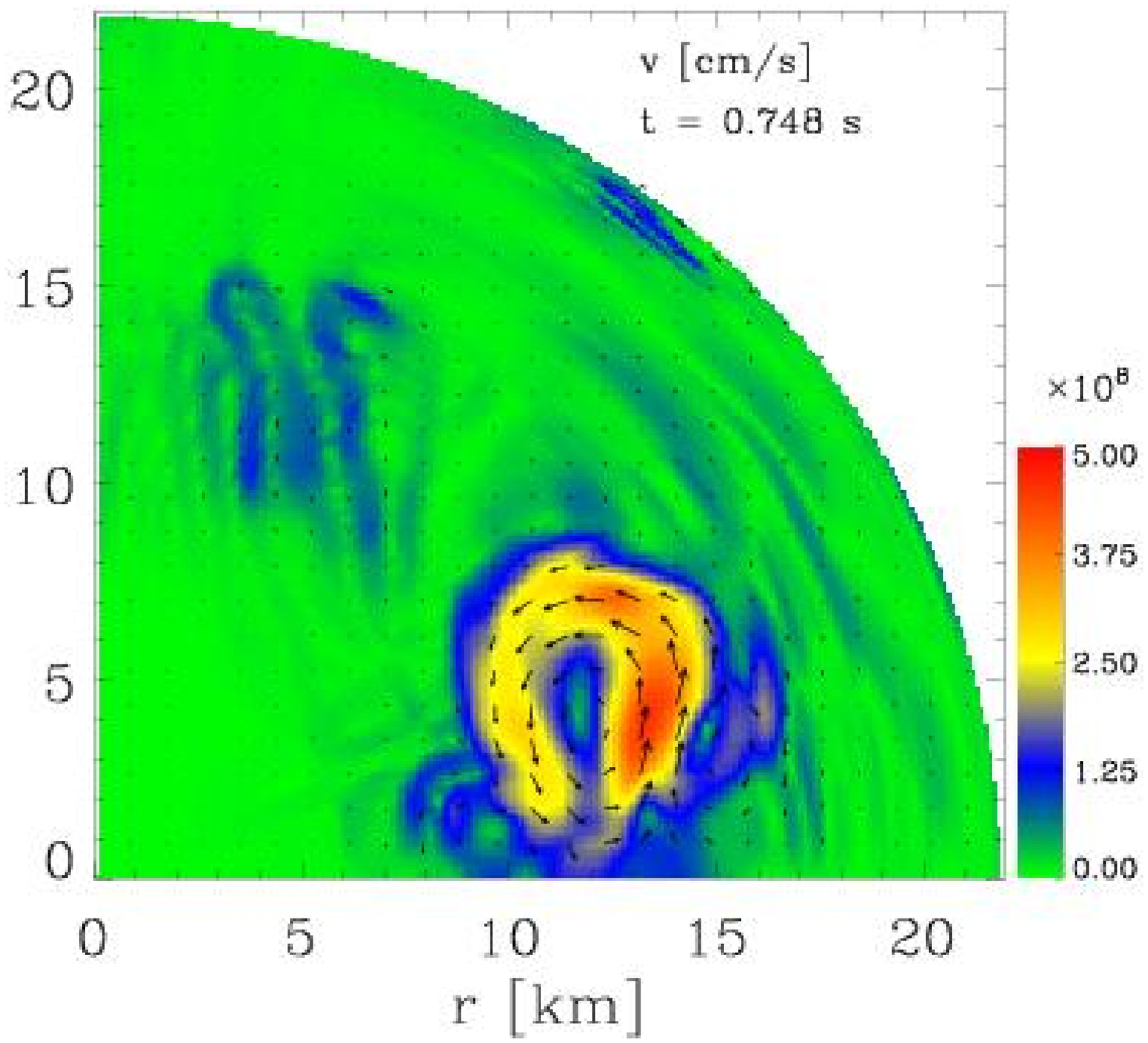,width=0.49\textwidth}}
\caption{\small Meridional cuts through a differentially rotating, 
         convective 
         nascent neutron star 0.75 seconds after its formation.
         The rotation axis coincides with the ordinates of the plots.
         In a quasi-steady state, convection is strongest at  
         intermediate radii in a region of essentially constant 
         specific angular momentum near the equator.
         It is only weakly developed closer to the rotation axis
         where elongated, convective cells occur that are
         aligned parallel to the axis. A steep gradient
         of the specific angular momentum suppresses convective
         motion perpendicular to the axis in this region.
         In contrast, in a non-rotating star convective activity
         takes place in a spherical shell. The hydrodynamic
         simulation was carried out with neutrino diffusion
         taken into account
         (Keil 1997, Janka \& Keil 1998, Janka et al.\ 2001).
         {\em Top left:} Contours of constant density
         (solid lines) between
         $3.67\times 10^{10}\,$g$\,$cm$^{-3}$ and
         $2.68\times 10^{14}\,$g$\,$cm$^{-3}$, increasing with
         a factor of 1.37, and of constant temperature (dotted lines)
         between 4$\,$MeV and 30$\,$MeV with steps of 1$\,$MeV.
         {\em Top right:} Lepton fraction $Y_{\mathrm{l}}$
         (density of electrons
         plus electron neutrinos minus their antiparticles
         relative to the number density of nucleons).
         {\em Bottom left:} Specific angular momentum $j_z$;
         the rotation period is $\sim$1$\,$ms at 3$\,$km
         distance from the rotation ($z$) axis, $\sim$2.5$\,$ms
         at 10$\,$km and $\sim$6$\,$ms at 20$\,$km.
         {\em Bottom right:} Total velocity in radial and
         lateral directions. The arrows indicate the flow
         direction in a meridional plane.}
\label{janka_fig1}
\end{figure}

The birth of a neutron star constitutes the transition of matter
on a macroscopic scale to the most extreme state realized after the
big bang. The throes are signaled by the conversion of huge amounts
of gravitational binding energy mostly to neutrinos
(up to $\sim$99\% or  
several 10$^{53}\,$erg), some to kinetic energy of the explosion
ejecta or wind loss ($\sim$1\%), and minor parts to
electromagnetic radiation ($\sim$10$^{49}\,$erg) and gravitational
waves ($\sim$10$^{46}\,$erg, possibly more).
These forms of energy release suggest potential observability,
but the rate of nearby SNe is rather low and available
empirical data are correspondingly sparse. Much of our
knowledge of neutron star formation and birth properties
like mass distribution, spins, magnetic fields, proper motions,
is therefore based on theoretical work, which, however, is
hampered by the enormous complexity of the problem and
barely constrained degrees of freedom, e.g.\ in the initial
conditions or input physics for models.
A brief moment in evolution therefore poses a big challenge
for exploration.

\section{Unveiling the Invisible: Signals from Birth}

Neutron stars are born as hot objects which initially 
contain a large number of electrons, electron neutrinos and 
protons. They lose their lepton content,
neutronize, and cool by the emission of neutrinos on a timescale
of several ten seconds (Burrows \& Lattimer 1986, Keil \& Janka 1995,
Pons et al.~1999). The neutrino burst from this phase was 
detected in case of Supernova~1987A in form of two dozen 
events by three underground experiments,
which thus opened the door to extragalactic neutrino
astronomy. The measurement of such a neutrino burst from a 
Galactic supernova with present experimental facilities could
yield tens of thousands of counts (e.g., Dighe, Keil, \& Raffelt 
2003), providing us with important information about the supernova
dynamics and neutron star equation of state 
(e.g., Pons et al.\ 2001).

\begin{figure}[htp!]
\centerline{\psfig{file=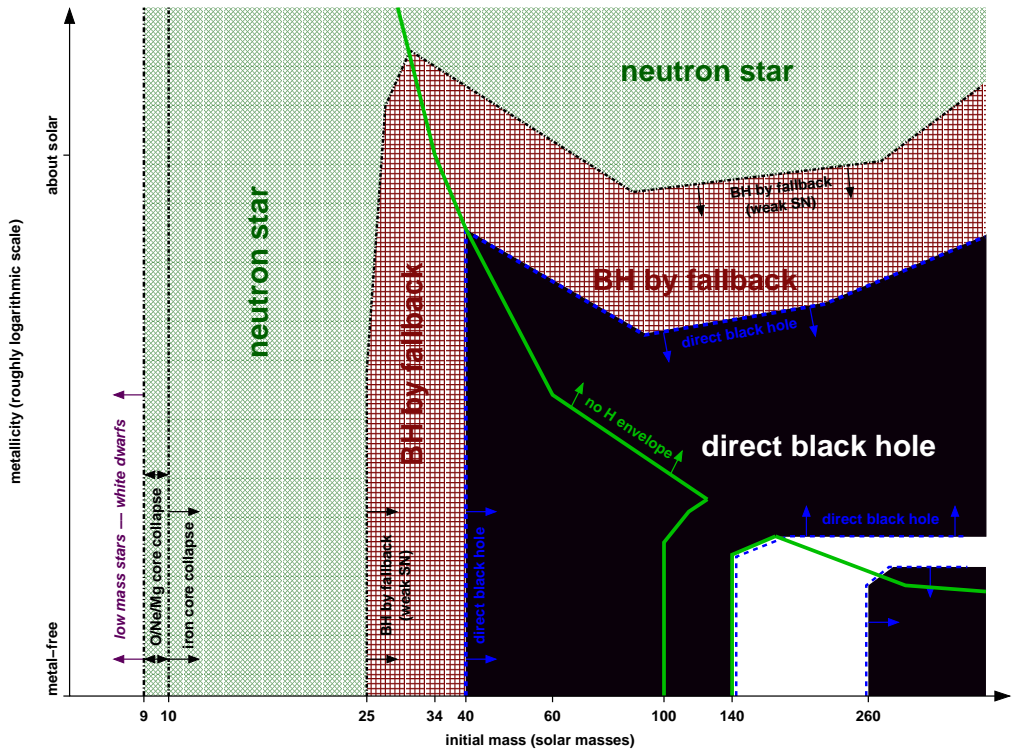,width=\textwidth,angle=270}}
\caption{\small Remnants of massive single (nonrotating) stars
         as a function of initial  
         metallicity. The evolution of a star depends on its mass
         loss, which increases for more massive stars and for higher
         content of metals in the stellar plasma. This explains
         the downward slope of the thick solid line above which
         the stars lose their hydrogen envelope.
         In the white strip near the
         right lower corner pair-instability supernovae leave no
         compact remnant, in the white region on the left side
         low-mass stars end their lives as white dwarfs
         (figure from Heger et al.\ 2003a).}
\label{janka_fig2}
\end{figure}

Similarly exiting would be the detection of gravitational
waves, which can be produced at the moment of core bounce
and by long-time post-bounce pulsations and oscillations, 
``r-mode instability'', or convective activity
in the interior of the newly formed compact remnant (e.g.,
M\"uller et al.\ 2004).
Immediately after stellar core collapse the
nascent neutron star develops convection in a region below the
neutrinosphere (e.g., Buras et al.~2003), driven mainly by a 
negative gradient of the lepton fraction, i.e, of the number
of leptons per baryon (Epstein 1979). This convective region 
digs deeper into the neutron star and encompasses an
increasingly thicker spherical shell as time advances.
Two-dimensional hydrodynamic simulations including neutrino 
diffusion (Keil 1997; Keil, Janka, \& M\"uller 1996) were able
to follow the evolution of the contracting, convective
proto-neutron star over a period of more than one second and
confirmed the picture suggested by stability analysis
of spherically symmetric models (Burrows 1987; Miralles, 
Pons, \& Urpin 2000). Self-consistent simulations with
rotation showed that the angular momentum transport by 
convection quickly (within only $\sim$100$\,$ms) leads to 
a quasi-steady state with highly differential rotation 
in which convection is strong near the equatorial plane 
but suppressed near the poles and close to the rotation
axis. This suppression is caused by the stabilizing effect
of a steep increase of the specific angular momentum $j_z$
with distance from the axis in the corresponding region
(Fig.~\ref{janka_fig1}; 
Janka \& Keil 1998; Janka, Kifonidis, \& Rampp 2001).

How frequent are the events which produce such powerful 
neutrino and gravitational wave signals? What are their
progenitors? And what can be said about the characteristic
properties of the forming neutron stars?

\section{Progenitors}

Recent counts of SN rates in galaxies of different morphological
types suggest that Type II and Ib,c SNe from stellar core-collapse 
events happen in our Galaxy (Sb-Sbc) with a rate of $1.5\pm 1.0$ per 
century (Cappellaro \& Turatto 2000, Cappellaro, Evans, \& Turatto 1999).
This means that roughly once every $\sim$65 years a compact remnant
--- a neutron star or black hole ---
should be formed. This estimate is a factor of two lower than values
based on recorded historical SNe or SN remnants (e.g., Strom 1994), 
but considering the large uncertainties both numbers are in
reasonable agreement. Another uncertainty, which, however,  
enlarges the error bars only insignificantly, is associated 
with the unknown rate of faint events, stellar core collapses 
which do not produce bright supernovae and thus could only be 
discovered by neutrino or gravitational wave measurements.

Accretion induced collapse (AIC) of white dwarfs in binaries,
which was invoked to explain large populations of millisecond
pulsars in globular clusters (Grindlay \& Bailyn 1988),
is likely to contribute only at a minor level.
Fryer et al.~(1999) constrained the possible rate
of such dim neutron star formation events by observed Galactic
element abundances, which limit the integral amount of 
ejected neutron-rich
matter and associated production of particular isotopes
(e.g., $^{62}$Ni, $^{66}$Zn, $^{68}$Zn, $^{87}$Rb, and $^{88}$Sr).
From their collapse models they deduced an
allowed event rate of typically a few $\sim$10$^{-5}$ per year
and at most several $\sim$10$^{-4}$ per year. This is two to
three orders of magnitude lower than the stellar core
collapse rate (see also Woosley \& Baron 1992) and in rough
agreement with the AIC rate needed to account for 
millisecond pulsars in globular clusters (Bailyn \& Grindlay 1990).
This agreement of both limits, however, may be purely accidentally,
because the amount of n-rich ejecta could be much smaller than
predicted by the simulations. It is sensitive to the neutron and
proton interactions of $\nu_e$ and $\bar\nu_e$ that are radiated
from the newly formed neutron star. The relevant physics is 
described only rather approximately in the existing simulations.

\begin{figure}[htp!]
\centerline{\psfig{file=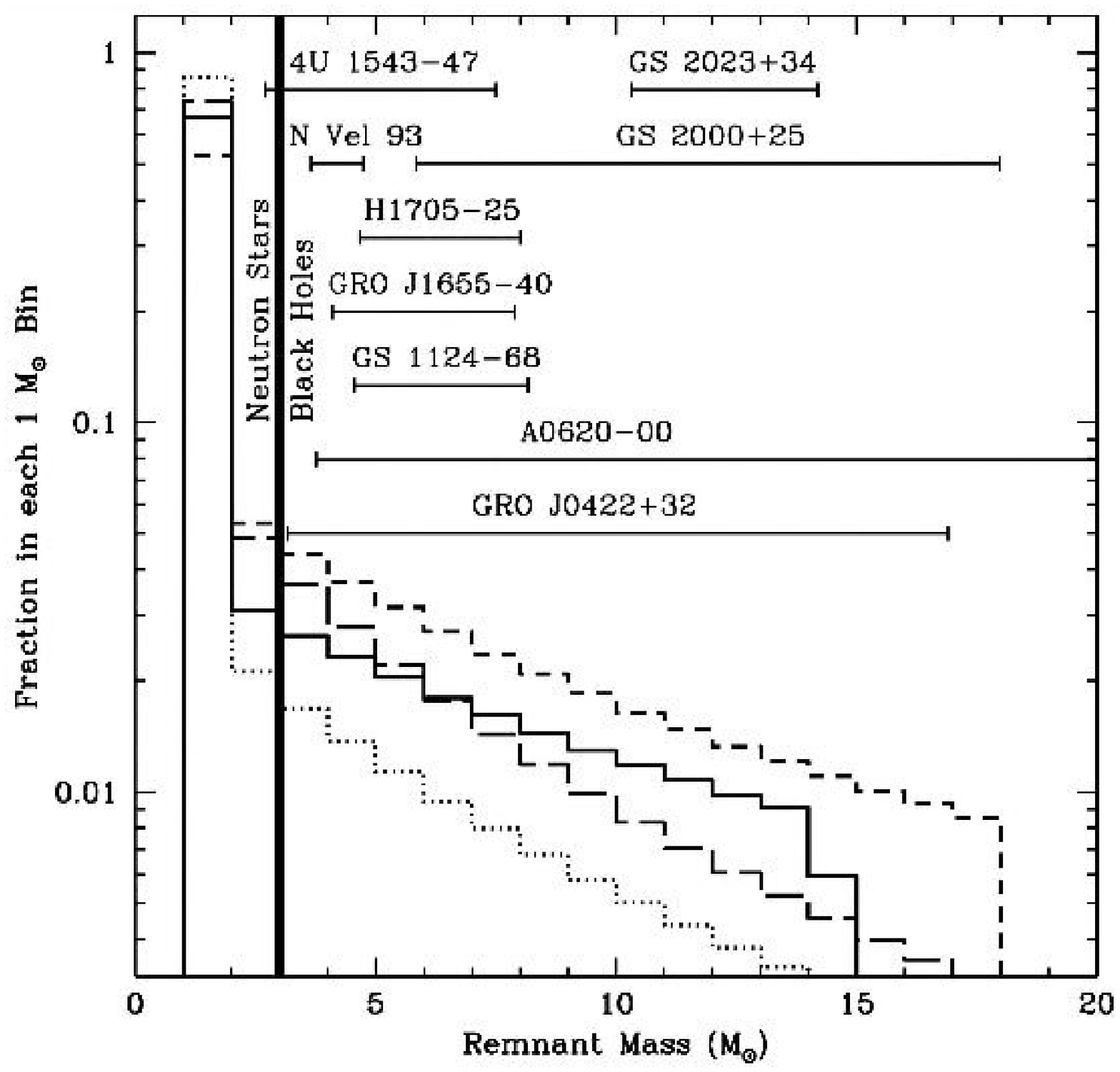,width=0.60\textwidth}
            \psfig{file=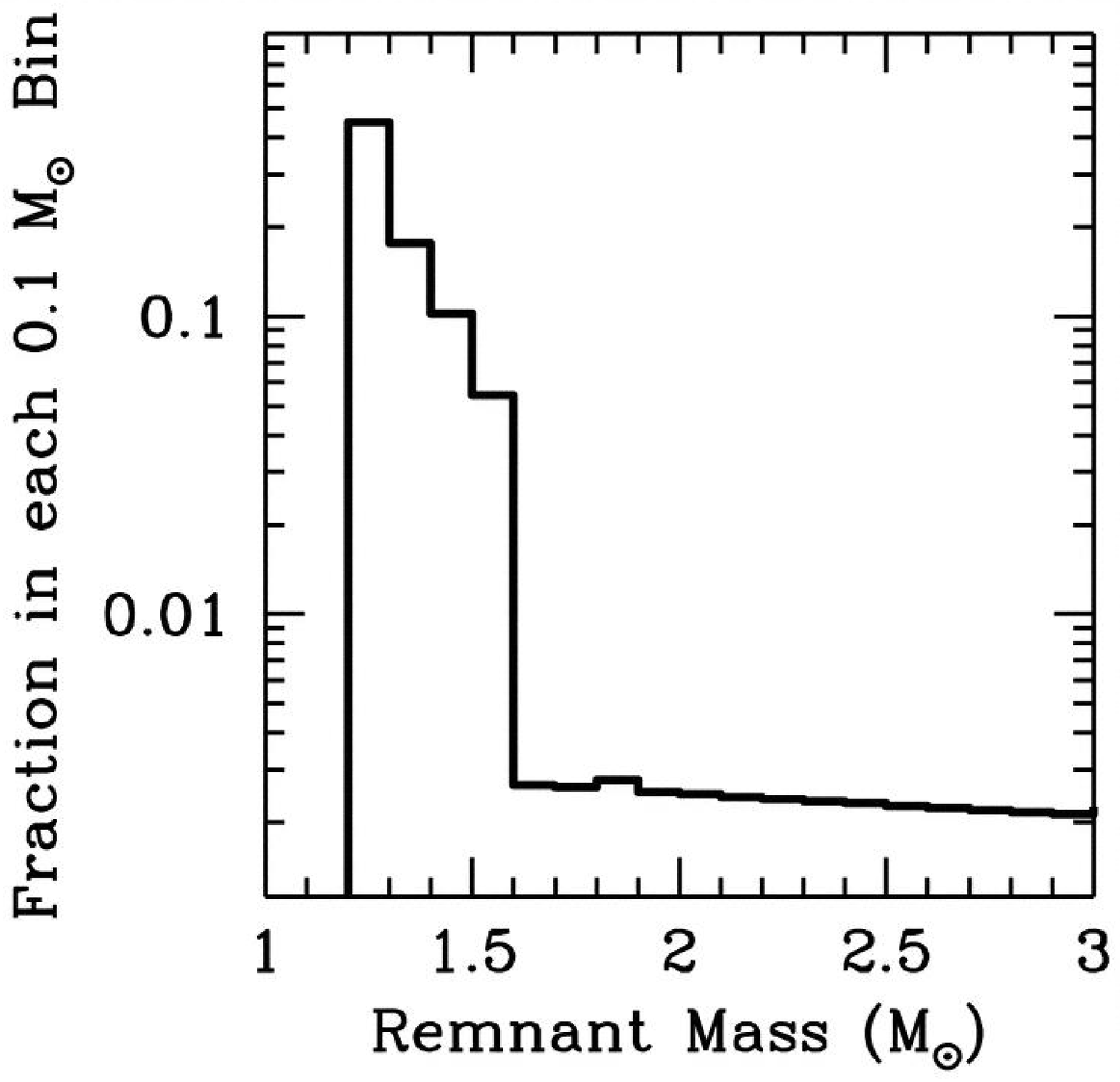,width=0.38\textwidth}}
\caption{\small {\em Left:} Mass distribution of compact remnants
         predicted from supernova simulations by Fryer. The
         different curves use different stellar initial
         mass functions and different estimates of the amount
         of fall back in the explosion. 60--80\% of the
         compact remnants are neutron stars if black holes
         are formed above $\sim$3$\,M_{\odot}$.
         {\em Right:} Calculated distribution of neutron-star
         remnant masses. Both figures are taken from
         Fryer \& Kalogera (2001).}
\label{janka_fig3}
\end{figure}

Stellar core collapse can lead to the formation of
a neutron star or a black hole, depending on the initial mass
and mass loss of the star during its evolution, both of which
determine the core mass before collapse. Moreover, the initial
metallicity of the star has crucial influence, since the mass
loss in stellar winds is sensitive to the opacity and thus
metallicity of the stellar gas. Recent reviews of the evolution 
of single stars and their final stages were given by 
Heger et al.~(2003a) and Woosley, Heger, \& Weaver~(2002). 
Figure~\ref{janka_fig2} (taken from Heger et al.~2003a) shows
that for solar metallicity neutron stars are expected to emerge
from stars above $\sim$9$\,M_{\odot}$ and below about
$25\,M_{\odot}$. Above this mass black holes form either
by fall back of matter which is unable to become unbound in the
supernova explosion or directly if the stellar core is too
massive to allow for the launch of an outward moving supernova 
shock. Fryer (1999) determines the limit for direct black hole 
formation to be around $40\,M_{\odot}$ and for black hole formation
by fall back to be somewhere between $18\,M_{\odot}$ and 
$25\,M_{\odot}$. The uncertainty of the latter limit is associated
with the progenitor structure in this stellar mass range and with
yet unresolved problems of the core-collapse physics which 
prohibit definite predictions of the supernova explosion energy.
Above about $33\,M_{\odot}$ the stars lose their whole hydrogen
envelope before collapse and become Wolf-Rayet stars with strong
mass loss (Woosley et al.~2002). For a sufficiently high mass loss
during this phase a ``window'' may exist above $\sim$50$\,M_{\odot}$
where again neutron stars are formed (Fig.~\ref{janka_fig1}).

\section{Masses}

Theoretical neutron star and black hole mass distributions 
were determined through
supernova simulations by Fryer \& Kalogera (2001).
Figure~\ref{janka_fig3}, which is taken from the latter paper,
shows that neutron stars are dominantly produced in the 
mass range 1.2--$1.6\,M_{\odot}$, above which the distribution of 
compact remnant masses falls off exponentially. 
With an adoped maximum neutron star mass of $\sim$3$\,M_{\odot}$
about 80\% of the compact remnants are neutron stars. 
The results show some variation with different assumptions
about the initial mass function of progenitors, supernova
energies, initial mass of the protoneutron star, and amount 
of fall back, as well as binary effects.  

A close comparison with measured neutron star masses is also
hampered by the small number of accurate mass measurements
on the one hand and the unclear exact relation between
(observable) gravitational masses and (computed) baryonic masses
on the other. This mass relation depends on the properties of the
high-density equation of state which determines the mass to radius
ratio and the binding energy of the neutron star. But current
theoretical analysis at least does not reveal any obvious
inconsistency with known masses of neutron stars (Thorsett \&
Chakrabarty 1999) and black hole candidates (Bailyn et al.~1998).

The nature of the compact remnant in Supernova~1987A is still
unclear. The detection of two dozen neutrinos from SN~1987A
was an unambiguous signal of the formation of a protoneutron star.
This star, however, could have become gravitationally unstable after
seconds of neutrino cooling, either triggered by the accretion 
of fall back matter (e.g., Brown, Bruenn, \& Wheeler 1992) or 
by a gradual softening of the supranuclear equation of state 
due to phase transitions
(e.g., Brown \& Bethe 1994, Keil \& Janka~1995, Glendenning 1995).
The existence of a black hole in SN~1987A is therefore not
excluded, but there is also no compelling argument in favor of
this possibility. If the neutron star has either a low magnetic
field or a low rotational frequency, it could well
remain invisible (Fryer, Colgate, \& Pinto 1999) just like
the faint compact object in Cas~A has been until a few years ago.

\section{Spins and Magnetic Fields}

Little is known about the rotation rates and magnetic fields
of newly born neutron stars. Only recently has stellar evolution
theory begun to include the transport of angular momentum  
in models which attempt to follow massive 
star evolution up to core collapse 
(Heger, Langer, \& Woosley 2000, Heger et al.\ 2003b). 

Massive stars are seen to rotate rapidly at the surface, but their
core properties need to be determined by numerical calculations.
During stellar evolution angular momentum is lost through
mass outflow in winds, and transported from the interior to the
surface by convection, shear, circulation, and magnetic torques.

Stars that start their evolution on the zero-age main sequence
with characteristic equatorial rotation velocities around
$\sim$200$\,$km$\,$s$^{-1}$ at the surface (corresponding to about
10\% of the break-up velocity) retain a specific angular momentum
of typically some $10^{16}\,$cm$^2$s$^{-1}$ prior to collapse. 
This is so large that neutron stars spinning with sub-millisecond 
periods would emerge. Such rotation near the critical limit is 
associated with a hugh rotational energy (several $10^{52}\,$erg)
which would be set free when the neutron star is 
decelerated to measured periods of young pulsars 
($\ga 10\,$ms; Marshall et al.\ 1998). There is no observational
evidence of any such gigantic energy release from ordinary
supernovae. On the other hand, specific angular momenta in excess
of $10^{16}\,$cm$^2$s$^{-1}$ just outside the core
are needed for progenitors of gamma-ray bursts (e.g., 
Woosley \& Heger 2003).

Including the effects of magnetic torques during stellar
evolution reduces the core angular momentum by roughly a 
factor of 20, which suggests the formation of young
neutron stars with periods of several milliseconds 
(Heger et al.\ 2003b). This appears to be in reasonable agreement
with observational constraints, although some slowing by 
neutrino-powered magnetic winds in case of magnetar-strength 
($\sim$10$^{15}$G) ordered surface fields 
(Thompson, Chang, \& Quataert 2004), the propeller mechanism 
in case of fall back of slowly moving supernova ejecta, or the
pulsar radiation mechanism might take place during the first 
seconds, days, or years, respectively, of the neutron star's
life (Woosley \& Heger 2003).

Pre-collapse magnetic fields are estimated to be
around $5\times 10^9\,$G for the toroidal component and
of order $10^6\,$G for the radial part (Heger et al.\ 2003b).
The field strength can increase during core collapse by a
factor of $\sim$1000. Further field amplification on 
timescales of hundreds of milliseconds to seconds could 
occur due to differential rotation by the magneto-rotational
instability (Akiyama et al.\ 2003), field winding (e.g.,
M\"uller \& Hillebrandt 1979), or due to convection by
dynamo action (Thompson \& Duncan 1993). Realistic and
quantitatively meaningful simulations have still to be done. 

It is pointed out here that neutrinos, although carrying away
a fair fraction of the rest-mass energy of the forming 
neutron star ($\sim$0.16$\,M_{\odot}c^2
(M_{\mathrm{ns}}/1.4\,M_{\odot})^2$ 
$=0.12\,M_{\mathrm{ns}}c^2(M_{\mathrm{ns}}/1.4\,M_{\odot})$;
Lattimer \& Prakash 2001),
are rather inefficient in removing angular momentum. 
The total angular momentum, $J_{\mathrm{ns}}$, changes when
the (gravitational) mass of the neutron star is reduced by
neutrino emission according to
(Epstein 1978, Baumgarte \& Shapiro 1998):
\begin{equation}
{{\dot J_{\mathrm{ns}}}\over J_{\mathrm{ns}}}\,=\,
            q\,{f(\lambda)\over \kappa_n}\,
            {{\dot M_{\mathrm{ns}}}\over M_{\mathrm{ns}}} \,,
\label{janka_eq1}
\end{equation}
where $q$ is an efficiency parameter ($q = 0$ if the neutrinos 
are emitted from the center, $q = 1$ for homogeneous volume 
emission without scattering, $q = 5/3$ for the diffusion case).
The function $f(\lambda)$ accounts for the deformation of the
rotating star and depends on the ratio of polar to equatorial
radius, $\lambda = (R_{\mathrm{p}}/R_{\mathrm{e}})^2$. 
Assuming ellipsoidal shape, it is
$f(\lambda) = 0.60(1+4\lambda)/(1+2\lambda)$. Note that 
the angular momentum loss relative to the energy loss 
decreases when the object becomes more oblate:
$f = 1$ for $\lambda = 1$ whereas $f = 0.6$ for $\lambda = 0$.
The factor $\kappa_n$
is a dimensionless structure constant of order unity which
depends on the density profile of the star. For a polytropic
equation of state, $P = K\rho^{1+1/n}$, 
$\kappa_n$ can be derived from the 
Lane-Emden function, which gives, e.g., 
$\kappa_n = 1$ for $n = 0$, $\kappa_n = 0.81482$
for $n = 0.5$, $\kappa_n = 0.65345$ for $n = 1$
(cf.\ Table~1 in Lai, Rasio, \& Shapiro 1993). 
Taking $qf(\lambda)/\kappa_n \approx\,$const during the 
cooling and contraction of the neutron star, 
Eq.~(\ref{janka_eq1}) can easily be integrated to yield
the ratio of final to initial quantities:
\begin{equation}
{J_{\mathrm{ns}}^{\mathrm{f}}\over J_{\mathrm{ns}}^{\mathrm{i}}}
\,=\,
\left({M_{\mathrm{ns}}^{\mathrm{f}}\over 
M_{\mathrm{ns}}^{\mathrm{i}}}\right)^{qf(\lambda)/\kappa_n} \,.
\label{janka_eq2}
\end{equation}
Pushing all involved numbers to their extrema, i.e.,
$q = 5/3$, $1/\kappa_n = 1.53$, $f(\lambda) = 1$, and 
$M_{\mathrm{ns}}^{\mathrm{f}}/M_{\mathrm{ns}}^{\mathrm{i}} = 0.8$,
we get $J_{\mathrm{ns}}^{\mathrm{f}}/J_{\mathrm{ns}}^{\mathrm{i}} 
= 0.57$ and for the specific angular momenta, $j = J/M$:
$j_{\mathrm{ns}}^{\mathrm{f}}/j_{\mathrm{ns}}^{\mathrm{i}}
= 0.71$. This means that neutrinos can remove at most 43\%
of the total angular momentum and thus reduce the specific
angular momentum by $\sim$30\% relative to its initial value.

\begin{figure}
\centerline{\psfig{file=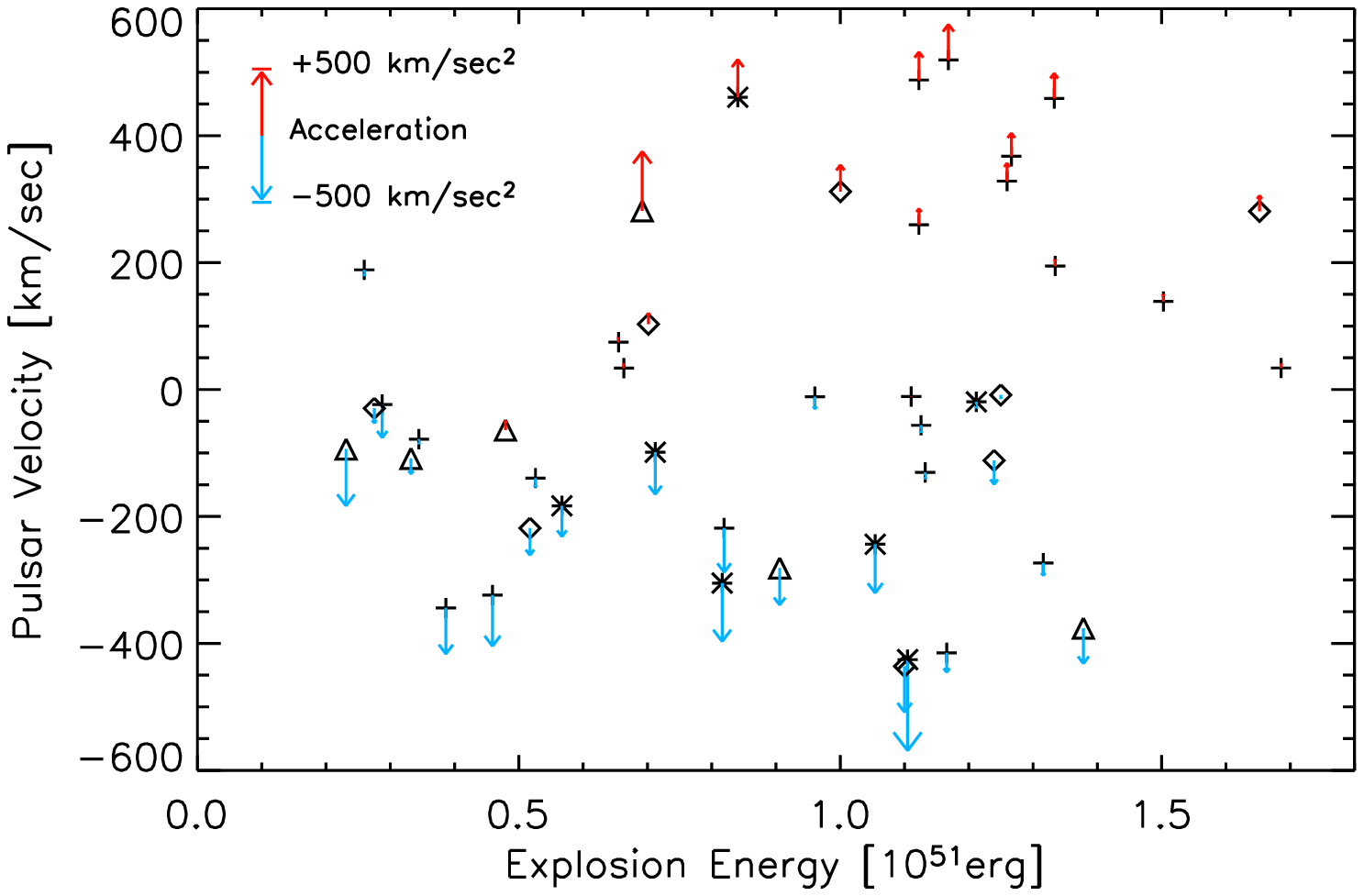,width=0.63\textwidth,angle=90}
            \psfig{file=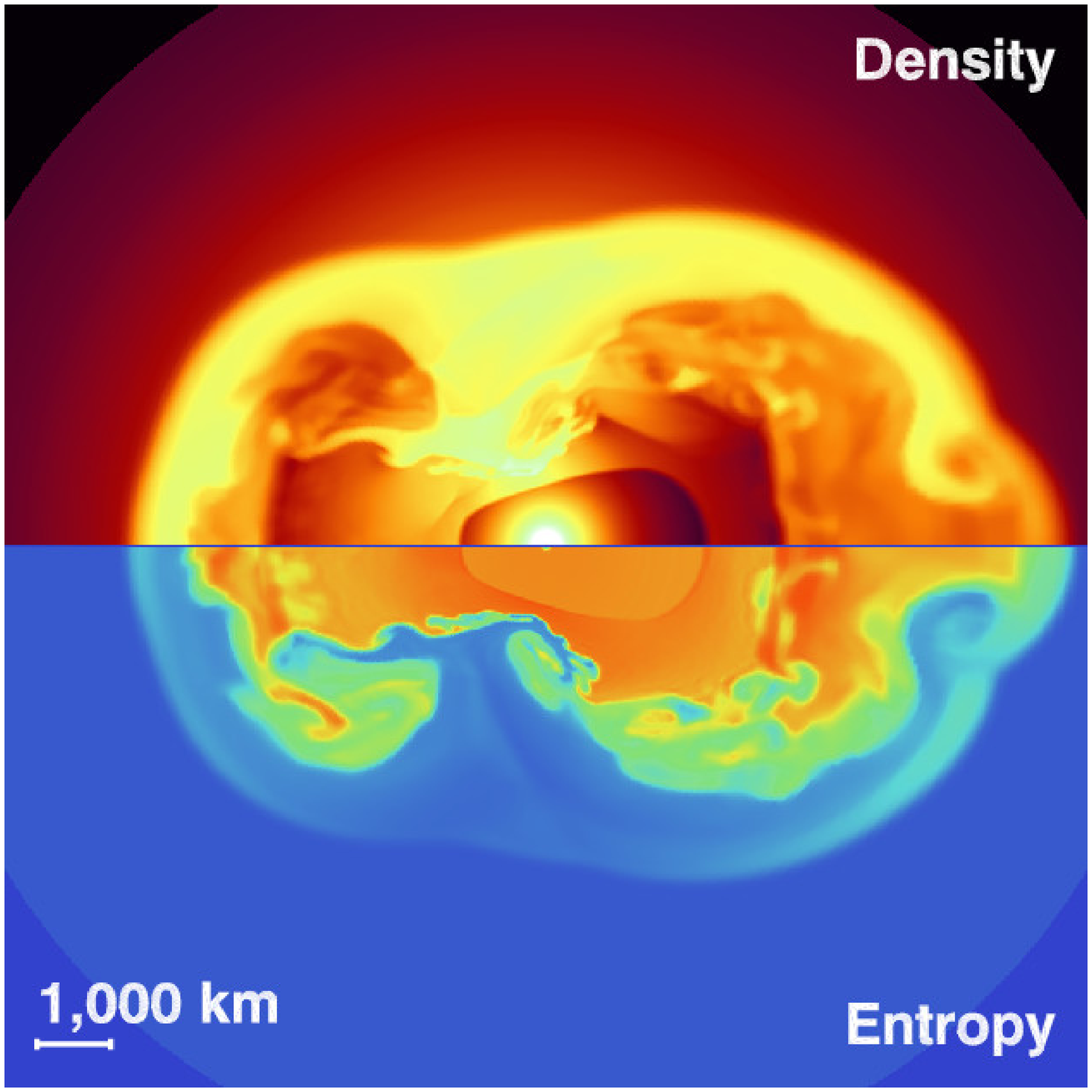,width=0.37\textwidth}}
\caption{\small {\em Left:} Kick velocities of pulsars versus 
         supernova explosion energy after one second of shock
         evolution as obtained in a sample of about 50 simulations.
         The length of the arrows indicates the size of the
         continuing acceleration at one second post bounce 
         when the simulations were stopped. Positive
         and negative values correspond to pulsar
         kicks in both directions of the symmetry axis 
         of the two-dimensional models (Scheck et al.\ 2004).
         The pulsar receives the opposite momentum of the
         explosion ejecta which can develop large global asymmetry 
         due to the non-linear growth of hydrodynamic instabilities
         in the neutrino-heating region between neutron star
         and shock. 
         {\em Right:} One of the models, one second after stellar
         core collapse. Density (top) and entropy distribution
         are shown. The explosion has an energy of 
         $1.4\times 10^{51}\,$erg and is stronger towards the
         right direction. The neutron star (invisible at the 
         center) correspondingly receives a recoil velocity of
         380$\,$km/s to the left.}
\label{janka_fig4}
\end{figure}

\section{Pulsar Kicks}

Young neutron stars are observed to possess average space 
velocities of 200--500$\,$km$\,$s$^{-1}$ (Lyne \& Lorimer 1994),
a significant fraction might move even faster than 
1000$\,$km$\,$s$^{-1}$
(Arzoumanian, Chernoff, \& Cordes 2002). These velocities are most
likely imparted to the neutron star by a kick associated with
its birth. Binary breakup is not sufficient to account for the
measured velocities. An intrinsic acceleration is also
required to explain special properties of neutron star binaries
(for reviews, see Lai, Chernoff, \& Cordes 2001 and Lai 2001).

Suggested mechanisms for natal neutron stars kicks can 
be grouped into two categories. Either they ascribe the pulsar
velocities to anisotropies of the supernova explosion or they
attribute them to the recoil associated with anisotropic neutrino 
emission during the neutrino cooling phase. A global asymmetry of
the radiated neutrinos of only $\sim$3\% yields 
$v_{\mathrm{ns}}\approx 1000\,$km$\,$s$^{-1}(E_\nu/3\times 10^{53}
{\mathrm{erg}})(M_{\mathrm{ns}}/1.4\,M_{\odot})^{-1}$.
However, producing asymmetries in the protoneutron star interior
which are sufficiently large to account for even only 1\% emission
anisotropy turns out to be extremely difficult and requires
ultrastrong magnetic fields ($\sim$10$^{16}$G) and/or
speculative assumptions about neutrino properties (e.g., 
Lai et al.\ 2001, Nardi \& Zuluaga 2001).

There exists, on the other hand, firm observational evidence
for large asymmetries in supernovae (e.g., inferred
from visible deformation of the ejecta or polarization 
measurements), and there is general agreement that hydrodynamic
instabilities lead to large-scale overturn and mixing 
behind the supernova shock already
during the very early moments of the explosion 
(e.g., Janka et al.\ 2003). Recently it was shown by 
simulations that such instabilities can produce global 
explosion aymmetries by which a net impulse of more than
500$\,$km$\,$s$^{-1}$ can be transferred to the neutron star
on a timescale of one second or longer 
(Fig.~\ref{janka_fig4}; Scheck et al.\ 2004;
see also Thompson 2000). Alternatively, inhomogeneities in
the core of the evolved star might grow during core collapse
(Lai 2001 and references therein) and might cause an anisotropic
shock breakout and large pulsar recoil (Burrows \& Hayes 1996;
but see Fryer 2004). If statistical hints of a two-component
velocity distribution of radio pulsar data (e.g., Arzoumanian et 
al. 2002) bear truth (and not just have the trivial meaning 
that more free parameters allow for better fits), different
mechanisms may be at work in subsets of the neutron star 
population.

{\small
\acknowledgments
Support by the Sonderforschungsbereich SFB-375
``Astro-Teil\-chen\-phy\-sik'' and the 
Sonderforschungsbereich-Transregio~7 
``Gravitationswellen-\-Astro\-no\-mie''
of the 
Deutsche Forschungsgemeinschaft 
is acknowledged.

}

\end{document}